\documentclass[runningheads,orivec]{llncs}

\usepackage[T1]{fontenc}
\usepackage[scaled=0.85]{beramono}

\usepackage{appendix}
\usepackage{cleveref}
\usepackage{listings}
\usepackage{tikz}
\usepackage{xcolor}

\usetikzlibrary{arrows, calc, decorations.pathreplacing}

\title{Coroutines with Higher Order Functions}

\author{Dimitri Racordon\inst{1}}
\authorrunning{D. Racordon}
\institute{
  University of Geneva, Switzerland\\
  \email{dimitri.racordon@unige.ch}
}

\begin{document}
\maketitle

\begin{abstract}
Coroutines are non-preemptive concurrent subroutines that,
unlike preemptive threads,
voluntarily transfer control between each others.
Introduced in the 60s before loosing in popularity in the 80s,
they have seen a regain of interest in recent years,
thanks to how elegantly they can solve numerous algorithmic problems.
Unfortunately, some mainstream languages still lack support for coroutines,
hence requiring either the use of non-standard interpreter/compilers,
or elaborate hacks in thrid-party libraries.
In this short paper,
we propose a very simple way to implement coroutine-like components
on the top of any language that support or can emulate higher order functions.
We accompany our explanations with a handful of examples in JavaScript.
\end{abstract}


\definecolor{bgdcolor} {RGB} {245, 245, 245}
\definecolor{cmtcolor} {RGB} { 81,  99, 116}
\definecolor{litcolor} {RGB} {  0, 128,   0}

\definecolor{kwcolor1} {RGB} {180,  88, 172}
\definecolor{kwcolor2} {RGB} {180,  88,   0}
\definecolor{kwcolor3} {RGB} { 77,  77, 255}


\lstset{
  captionpos        = b,                   
  extendedchars     = true,                
  tabsize           = 2,                   
  columns           = fixed,               
  keepspaces        = true,                
  showstringspaces  = false,               
  breaklines        = true,                
  numbers           = left,                
  basicstyle        = \small\ttfamily,     
  numberstyle       = \tiny\ttfamily,      
  commentstyle      = \color{cmtcolor},    
  stringstyle       = \color{litcolor},    
  keywordstyle      = [1]\color{kwcolor1},
  keywordstyle      = [2]\color{kwcolor2},
  keywordstyle      = [3]\color{kwcolor3},
  belowskip         = -0.8 \baselineskip,  
  belowcaptionskip  = 0 \baselineskip,     
  xleftmargin       = 2em,                 
  framexleftmargin  = 2em,                 
  xrightmargin      = -2em,                
  framexrightmargin = -2em,                
}

\lstset{literate=%
   *{0}{{{\color{litcolor}0}}}1
    {1}{{{\color{litcolor}1}}}1
    {2}{{{\color{litcolor}2}}}1
    {3}{{{\color{litcolor}3}}}1
    {4}{{{\color{litcolor}4}}}1
    {5}{{{\color{litcolor}5}}}1
    {6}{{{\color{litcolor}6}}}1
    {7}{{{\color{litcolor}7}}}1
    {8}{{{\color{litcolor}8}}}1
    {9}{{{\color{litcolor}9}}}1
}

\lstdefinelanguage{javascript} {
  keywords    = [1]{function, let, const, for, while, switch, case, default, break, yield, return},
  keywords    = [2]{true, false, null},
  keywords    = [3]{console},
  morecomment = [l]{\#},
  morestring  = [b]",
}

\lstdefinelanguage{python} {
  keywords    = [1]{class, def, if, for, in, while, yield, return},
  keywords    = [2]{True},
  keywords    = [3]{next, print, range},
  morecomment = [l]{\#},
  morestring  = [b]",
}

\section{Introduction}

Coroutines are subroutines that can voluntarily suspend their execution to yield control to another subroutine.
They can elegantly implement inversion of control patterns,
such as iterators~\cite{DBLP:conf/popl/LiuKM06} and generators~\cite{DBLP:conf/ipps/MillsJ16},
and offer an interesting way to reason about asynchronous code~\cite{DBLP:conf/ecoop/BiermanRMMT12}.

Support for coroutines is unfortunately lacking in some mainstream programming languages,
such as Java or Swift,
without resorting to non-standard interpreter/compilers~\cite{DBLP:conf/pppj/StadlerWW10},
or elaborate hacks in third party libraries~\cite{book/Mukherjee15}.
In this paper, we present a simple technique to implement coroutines on the top of a programming language,
without any need for additional control flow operations (e.g. \lstinline{setjmp} and \lstinline{longjmp} in C/C++),
nor any modification of the compiler/interpreter of the language.
Instead, our approach is purely based on higher order functions, or emulation thereof.
\section{Coroutine Semantics}

Coroutines were first introduced in 1963 by Conway~\cite{DBLP:journals/cacm/Conway63},
who described them as autonomous modules that may communicate with each others.
They lost popularity during the 80s in both academics and industries,
when preemptive systems prevailed,
but have since them found their way back into mainstream programming languages such as
Lua, Python, Ruby, Kotlin and many others.
Coroutines are very similar to regular routines (a.k.a. functions).
They usually accept some arguments and produce some result.
The difference lies in the fact that coroutine may suspend (typically with a \lstinline{yield} statement)
and transfer control to another coroutine.
It may then get the control back to resume from the exact point it paused,
having it state preserved so the execution can continue normally until it once again yield control,
or fully terminates.

Although, the above definition only describes the core concept of coroutines,
and not their exact semantics.
In fact, there exist various implementations with vastly different features
(see~\cite{DBLP:journals/toplas/MouraI09} for an elaborate discussion).
One important distinction is whether coroutines are implemented \emph{symmetric} or \emph{semi-symmetric}.
In the former model, control can be explicitly passed from one coroutine to any other,
while in the latter this is restricted to direct parent/child in the call stack.
Symmetric coroutines offer a finer control flow management,
but their semi-symmteric counterpart is generally easier to use and reason about.
Note however that both have been shown to have the same expressiveness~\cite{DBLP:journals/toplas/MouraI09}.
Another important characteristic of a particular implementation is whether or not coroutines are \emph{stackfull}.
Such coroutines can be suspended from deeper in their call stack (i.e. inside nested functions calls),
while other may not.
Writing non-blocking programs using asynchronous operations without stackfullness
requires libraries to be structured accordingly.
Namely, each function must act as a generator,
and each call must explicitly pull results out of the nested ones~\cite{url/Nystorm}.
However, just as symmetric and semi-symmetric coroutines are equally expressive,
stackfull coroutines do not improve on the expressiveness of their stackless counterpart.
In this paper, we will focus on stackless asymmetric coroutines.

\tikzstyle{io}       = [rectangle, thick, minimum size=3mm, rounded corners, text=white]
\tikzstyle{block}    = [rectangle, thick, draw=black!75, minimum size=6mm]
\tikzstyle{looptest} = [circle, thick, draw=black!75, minimum size=6mm]

\begin{figure}[ht]
\begin{lstlisting}[
  language={javascript},
  escapechar = {!},
  caption={Fibonacci sequence generator in JavaScript.},
  label={lst:background/fib},
]
function * fib() {!\tikz[remember picture] \node [] (lst-top) {};!
 !\tikz[remember picture] \node [] (lst-2) {};!let a = 0
 !\tikz[remember picture] \node [] (lst-3) {};!let b = 1
 !\tikz[remember picture] \node [] (lst-4) {};!while (true) {
 !\tikz[remember picture] \node [] (lst-5) {};!  yield a
 !\tikz[remember picture] \node [] (lst-6) {};!  const c = a
    a = b
 !\tikz[remember picture] \node [] (lst-8) {};!  b = c + a
  }
}
  
const g = fib()
for (let i = 0; i < 10; ++i) {
  console.log(g.next())
}
\end{lstlisting}
  \centering
  \begin{tikzpicture}[remember picture, overlay, >=stealth', node distance=1.25cm]
    \node[io, fill=black!40!green] at ($(lst-top) + (5cm ,0)$) (start) {\ttfamily start};
    \node[block, below of=start] (1) {1};
    \node[looptest, below of=1] (2) {2};
    \node[block, below left of=2] (3) {3};
    \node[block, below of=3] (4) {4};
    \node[io, fill=black!40!red, below right of=2] (end) {\ttfamily end};

    \path (start) edge[->] (1)
          (1) edge[->] (2)
          (2) edge[->] node[left] {\ttfamily yes} (3)
          (3) edge[->] (4);
    \draw[->, rounded corners] (4) -| (2);
    \draw[->] (2) -- node[right] {\ttfamily no} (end);
    
    \draw [decorate,decoration={brace,amplitude=2pt,raise=4pt},yshift=0pt]
      ($(lst-2.north)+(2.5cm, 0.08cm)$) -- ($(lst-3.south)+(2.5cm, -0.08cm)$)
      node[midway, xshift=0.3cm] (blk-1) {};
    \draw[->, dashed] (1) -- (blk-1);
    \draw [decorate,decoration={brace,amplitude=2pt,raise=4pt},yshift=0pt]
      ($(lst-4.north)+(2.5cm, 0.08cm)$) -- ($(lst-4.south)+(2.5cm, -0.08cm)$)
      node[midway, xshift=0.3cm] (blk-2) {};
    \draw[->, dashed] (2) -- (blk-2);
    \draw [decorate,decoration={brace,amplitude=2pt,raise=4pt},yshift=0pt]
      ($(lst-5.north)+(2.5cm, 0.08cm)$) -- ($(lst-5.south)+(2.5cm, -0.08cm)$)
      node[midway, xshift=0.3cm] (blk-3) {};
    \draw[->, dashed] (3) -- (blk-3);
    \draw [decorate,decoration={brace,amplitude=2pt,raise=4pt},yshift=0pt]
      ($(lst-6.north)+(2.5cm, 0.08cm)$) -- ($(lst-8.south)+(2.5cm, -0.08cm)$)
      node[midway, xshift=0.3cm] (blk-4) {};
    \draw[->, dashed] (4) -- (blk-4);
  \end{tikzpicture}
\end{figure}

Listing \ref{lst:background/fib} illustrates the implementation of a fibonacci sequence generator
by the means of a coroutine in JavaScript.
The control flow graph on its right will be discussed later an can be ignored for the time being.
Rather than ``returning'' like a regular routine would,
this coroutine produces a value for its caller and suspends with the \lstinline{yield} statement at line 5.
Note that the coroutine is \emph{instantiated} at line 12,
but has not executed any instruction yet at this point.
Instead, this happens at line 13,
when its instance \lstinline{g} is resumed,
and so it executes until it reaches line 5,
produces the next number of the sequence or suspends again.

  
\section{Rewriting Coroutines}

In most interpreters,
a coroutine is essentially nothing more than a function with a pointer to a call frame
that stores the value of its local variables,
as well as an instruction pointer indicating the next statement to execute.
The idea of our approach is to reproduce this setup by the means of higher order functions,
so as to store the value of the local variables and that of the instruction pointer as part of a function closure.
An example is given in Listings \ref{lst:impl/simple-coro} and \ref{lst:impl/simple-rewrite},
respectively illustrating a simple coroutine and its rewritten form,
as a higher order function.
The argument the coroutine gets on the left
is captured in the closure of the function that is returned on the right,
and so is an additional \lstinline{inst} variable that acts as the instruction pointer.
Finally, the values the coroutine may \emph{receive} upon being resumed on the left
are represented as arguments of the function that is returned on the right.

\begin{figure}
\centering
\begin{minipage}{.40\textwidth}
\begin{lstlisting}[
  language={javascript},
  caption={A simple coroutine},
  label={lst:impl/simple-coro},
  showlines={true},
  numbers={none},
]



function * f(n) {
  const x = yield n
  yield n + x
}



\end{lstlisting}
\end{minipage}\hfill
\begin{minipage}{.50\textwidth}
\begin{lstlisting}[
  language={javascript},
  caption={Its rewritten form},
  label={lst:impl/simple-rewrite},
  numbers={none},
]
function f(n) {
  let inst = 1
  return function(x) {
    switch (inst) {
    case 1 : inst += 1; return n
    case 2 : inst += 1; return n + x
    default: break
    }
  }
}
\end{lstlisting}
\end{minipage}
\end{figure}

An interpreter is likely to consume a rather low-level language,
where instructions corresponds to a single statement,
which may not be the case at a higher abstraction level.
As a consequence, we need to refine our notion of instruction pointer.
Indeed, rather than considering each \emph{line},
it is better to reason about where control flow may be affected,
that is at branches (e.g. \lstinline{if}, \lstinline{while}, \dots) and \lstinline{yield} statements.
Determining these \emph{control flow jumps} can be done easily by constructing
the \emph{control flow graph}~\cite{DBLP:conf/csmr/MorettiCO01} of a the coroutine.
An example is given in Listing \ref{lst:background/fib},
where the control flow graph on the right is that of the generator.
Squares represent blocks of synchronous code,
unaffected by control flow,
while circles represent branches.
Notice that the \lstinline{yield} statement at line 5 is represented as its own block,
as it obviously represents a control flow jump.

Such control flow graph highlights which parts of the coroutine will constitute instruction blocks in the function we will return,
and how will the instruction pointer be updated.
The last step is to wrap the entire instruction selection (i.e. the \lstinline{switch} statement in our example) in a loop,
so as to handle blocks that do not terminate with a \lstinline{yield} statement.
Consider for instance a simple test,
which will determine the next instruction to execute,
as depicted by the node with the label ``2'' in our example.
Executing such block will update the instruction pointer,
but will not produce a return value,
and one should then go back to the instruction selection in order to jump to the appropriate instruction block.

\begin{lstlisting}[
  float={ht},
  language={javascript},
  caption={Rewritten form of the fibonacci sequence generator of Listing \ref{lst:background/fib}.},
  label={lst:impl/fib}
]
function fib() {
  let inst = 1; let a = null; let b = null
  return function() {
    while (true) {
      switch (inst) {
      case 1:
        inst = 2; a = 1; b = 2
      case 2:
        inst = 3; return a
      case 3:
        inst = 2; const c = a; a = b; b = c + a
      }
    }
  }
}
\end{lstlisting}

Note that a minor optimization can be performed.
Indeed, branches that can always go one direction can be merged with their successors,
and diamonds on the graph that neither contain a \lstinline{yield} statement nor the resumption of another coroutine can be squashed together.
As a result, Listing \ref{lst:impl/fib} is an acceptable rewriting
of the fibonacci generator presented in Listing \ref{lst:background/fib},
where the conditional branch has been removed.

\subsection{Emulating Higher Order Functions}

It is possible to eliminate all higher order function from a program by replacing them with a record
that contain their environment,
and some sort of pointer to a \emph{lifted} first order function~\cite{DBLP:journals/lisp/Reynolds98}.
Therefore, one may also rewrite the higher order function produced by our approach into a first order one,
so as to emulate coroutine in a language without any support for higher order.
An implementation example of the fibonacci generator in C is proposed in the appendix.

\section{Related Work}

Early works on coroutine implementations,
which heavily relied on continuations~\cite{DBLP:journals/cl/HaynesFW86},
already showed that coroutines could be implemented without any modification to the core language.
However, one challenge of these approaches is to adapt a usually statement-oriented style of coroutine
to more expression oriented programs,
while ours stays closer to the semantics of the target language.

More recent works aim at implementing coroutine more efficiently,
often by adding built-in support to the interpreter or virtual machine~\cite{DBLP:conf/pppj/StadlerWW10}.
These approaches typically rely on the creation of dedicated stacks to keep track of a coroutine state.
Although our method does not emphasis on performances,
all information about a particular coroutine are contained within a single function closure,
therefore not requiring additional data structures or bookkeeping.
\section{Conclusion}

We present a simple approach to implement coroutines in any language that supports higher order function.
Our technique does not require any modification to the host language or its interpreter/compiler.
By emulating higher order function,
we also show that our approach is even applicable to languages without first-class functions.

Future works include a performance analysis of our approach,
so as to determine the overhead of our instruction selection process,
compared to more elaborate techniques,
such as that mentioned in related work.
Another interesting axis would be to implement a tool that rewrites coroutines automatically from a superset of the target language.
This would offer a non-invasive approach to the addition of cooperative multitasking in languages that do not support them natively.

\bibliographystyle{splncs04}
\bibliography{references}

\newpage
\begin{appendix}
\section{Fibonacci Sequence Generator in C}

\begin{lstlisting}[language={c}]
// Emulation of first order functions.
typedef struct {
  void* env;
  void* (*fn)(void*);
} function_t;

void* apply(function_t closure) {
  return closure.fn(closure.env);
}

// Rewriting of the fibonacci sequence coroutine.
typedef struct  {
  int inst;
  int a;
  int b;
} fib_env;

void* fib_fo(void* e) {
  fib_env* fe = (fib_env*)(e);
  while (1) {
    switch (fe->inst) {
    case 1:
      fe->inst = 2; fe->a = 1; fe->b = 2;
    case 2:
      fe->inst = 3; return &(fe->a);
    case 3:
      fe->inst = 2; int c = fe->a; fe->a = fe->b; fe->b = c + fe->a;
    }
  }
}

function_t fib() {
  fib_env* env = (fib_env*)(malloc(sizeof(fib_env)));
  env->inst = 1;
  function_t closure = { env, &fib_fo };
  return closure;
}

// Example of invocation.
int main() {
  function_t g = fib();
  for (int i = 0; i < 10; ++i) {
    printf("%i\n", *(int*)(apply(g)));
  }
  free(g.env);
  return 0;
}
\end{lstlisting}
\end{appendix}

\end{document}